# Optimization of Existing Centroiding Algorithms for Shack Hartmann Sensor

Akondi Vyas[1,*], M B Roopashree[2], B R Prasad[3]

*Indian Institute of Astrophysics, Bangalore, India*

[1]vyas@iiap.res.in

[2]roopashree@iiap.res.in

[3]brp@iiap.res.in

Indian Institute of Science
Bangalore, India

**Abstract**

*Three centroiding techniques to estimate the position of the spots in a Shack Hartmann sensor: Normalized Centre of Gravity (CoG), Iteratively Weighted Centre of Gravity (IWCoG) and Intensity Weighted (IWC) centroiding are studied in comparison. The spot pattern at the focal plane of a Shack Hartmann sensor was simulated by including the effect of a background noise. We present the results of optimization of the performance of each of the centroiding techniques as a function of Signal to Noise Ratio (SNR) at different experimental conditions.*

## I. INTRODUCTION

In the midst of increasing attempts for big telescopes in large number, adaptive optics has captured international astronomical attention [1]. The wave-front sensors have to share a vital responsibility in the case of wide field of view atmospheric correction. The Shack Hartmann sensor is an array of lenslets that is placed in the path of the light beam. The sensor detects the local slopes of the wave-front in terms of the shifts in spots produced at the focal plane of the sensor and computationally reconstructs the wave-front. With increasing telescope size, the Shack Hartmann sensor resolution increases and this leads to more computational effort. The performance of a Shack Hartmann sensor is limited by centroiding accuracy, wave-front reconstruction ability and discretization error [2]. Wave-front reconstruction ability depends on the geometry used for wave-front estimation from slope measurements and includes the errors due to centroiding. Discretization adds error that cannot be controlled once the selection of the resolution of the Shack Hartmann sensor is made. Well known problems in astronomical adaptive optics are low light levels, background noise effects and the problem of differential correction that leads to smaller shifts in the spots. Effectively, the task is to sense the small shifts in low light level zone inflicted by unwanted background noise. Optimization of the performance of various centroiding algorithms at different Signal to Noise Ratios can improve the wave-front reconstruction accuracy [3].

The centroiding methods studied in this paper include normalized Centre of Gravity (CoG), Iteratively Weighted Centre of Gravity (IWCoG) and Intensity Weighted Centroiding (IWC). The CoG method uses the averaging formula, which is the ratio of sum of products of position coordinate and intensity at that point to the total intensity. WCoG uses additional information that the spot has a Gaussian spread and weights the intensity function with a Gaussian intensity distribution, effectively making an attempt to fit a Gaussian to the spot. In IWCoG method, the position of the Gaussian centre and the spread are iteratively corrected to go closer to the actual centroid [4].

Each of these techniques has their advantages and disadvantages which can be investigated to optimize the centroiding performance in the case of Shack Hartmann kind of a sensor. CoG method has advantage over other techniques in the absence of noise. In the presence of background noise the performance of CoG method will be degraded depending on noise distribution. If the background noise is uniform and the number of pixels under observation (sample size) is large then statistically, the centroid of the noise will be closer to the image centre and will produce smaller centroiding error. The IWCoG method has an advantage that it can detect the centroid position more accurately even in the presence of noise, but at the cost of increased computational time and iteration convergence problems. The optimum number of iterations changes with Signal to Noise Ratio (SNR), the spot size and the shift in the spot. IWC method with weighting function equal to the intensity distribution is mathematically the optimum weighting function that must be used to minimize the centroid estimation error, although the additional information about the shape of the spot gives better estimate in the case of IWCoG. If the shape of the spot is not retained as a near Gaussian, then IWCoG fails to accurately locate the centroid, in which case IWC plays an important role. Using the higher powers of intensity as weighting function may not point the actual centroid, but helpful in picking out the point with maximum intensity.

The state of the art large telescopes need adaptive optics systems with speeds better than a few milliseconds. This requires faster and efficient computation. In this paper we presented the results of the performance of the three centroiding methods described earlier at different noise levels. The optimization of the performance of the algorithms will lead to a more efficient, accurate and faster centroid estimate.

## II. BACKGROUND

Let $I(x,y)$ be the intensity function corresponding to a subaperture of a Shack Hartmann sensor at its focal plane. The CoG method defines the centroid position $(x_c, y_c)$ as,

$$(x_c, y_c) = \frac{\sum_{i,j} X_{ij} I_{ij}}{\sum_{i,j} I_{ij}}$$

In the case of the weighted centroiding, we weight the centroid estimate formula by a weighting function, $W(x,y)$.

$$(x_c, y_c) = \frac{\sum_{i,j} X_{ij} I_{ij} W_{ij}}{\sum_{i,j} I_{ij} W_{ij}}$$

In the case of the Shack Hartmann sensor, the shape of the spot can be assumed to be a 2D Gaussian function $W(x,y)$ about the actual centroid position $(x_c, y_c)$ and hence,

$$W(x,y) = \frac{1}{2\pi\sigma^2} \exp\left\{-\frac{(x-x_c)^2}{2\sigma^2} - \frac{(y-y_c)^2}{2\sigma^2}\right\}$$

where $\sigma$ defines the spread. In the weighted centroiding technique we take the advantage of the knowledge of the shape of the spot. This method is advantageous when the spot maintains a similar shape. In the iterative process, we do not know the actual position and spread of the spot in the first iteration, hence, we use image centre as the prior and the information of the centroid position and size obtained in the first iteration is fed to next iteration.

### A. Simulations

Simulation of the spot pattern at the focal plane of a Shack Hartmann sensor was done in two steps as follows:
- A 2D Gaussian intensity pattern was simulated with an image size of 50×50 pixels. The centre of the spot was positioned at a known point on the image. The shifts are measured with respect to the image centre.
- 2D spatially uniform noise was then superimposed whose intensity distribution function follows Gaussian statistics.

Fig. 1 shows the simulated spot of size 6 pixels shifted by 2 pixels at different SNR.

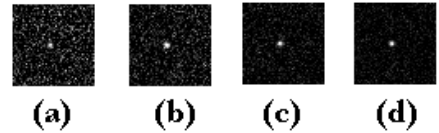

Fig. 1 Simulated spot at SNR (a) 0.3 (b) 0.5 (c) 0.7 and (d) 1.0

The performance of the centroiding algorithms was analysed using the percentage centroid estimation error (CEE) defined as shown below:

$$\text{CEE} = \frac{\sqrt{(x_c - x_c^*)^2 + (y_c - y_c^*)^2}}{S} \times 100$$

where ($x_c$, $y_c$) represents the position of the actual centroid, ($x_c^*$, $y_c^*$) is the estimated centroid position using the algorithms and S is the amplitude of shift of the spot from the image centre.

### III. COMPUTATIONAL STUDIES

The performance of the algorithms is presented in the subsections as a function of SNR at different conditions like the size of the spot, the shift in the spot. The question of optimum number of iterations for minimum centroiding error is addressed in the case of IWCoG. The effect of using higher powers of intensity as weighting function is analysed. The issues of computational speeds, error stability, temporal stability are addressed.

#### A. Performance of algorithms at different SNR:

The centroid estimation error is plotted as a function of SNR for each of the techniques discussed earlier and shown in Fig. 2. Average CEE represents the mean value of 1000 trials.

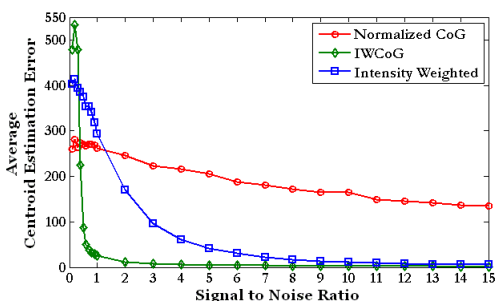

Fig. 2 Comparison of algorithms at different SNR, 0.2 pixel shift

All the curves cross each other below SNR=1. This computational experiment was done in the case of spot of size 6 pixels shifted from the centre by 0.2 pixels. Even IWCoG that has less CEE does not seem to be useful below SNR=0.5, because the error is large compared to the shift in the spot. IWC can be used if SNR > 3. Normalized CoG cannot be used for very small shift in the spots and low SNR. We can see the case of larger shift (8 pixels) in Fig. 3.

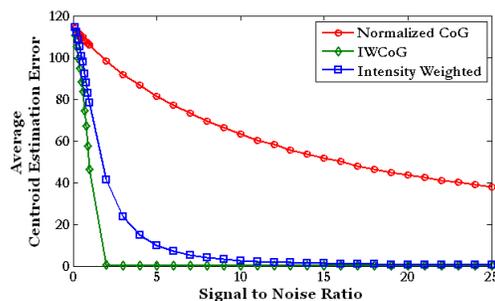

Fig. 3 Comparison of algorithms at different SNR, 8 pixel shift

Clearly, for larger shift in the spot, the CEE reduces for all the algorithms.

#### B. Dependence on the shift of the spots:

To understand the dependence of the centroiding accuracy on the amplitude of the shift in the spot from the image centre, we simulated images with shifts starting from 0.05 pixels to 1.5 pixels. The correlation between the amplitude of the shift and the centroiding accuracy for different methods is presented below:

1) *CoG*: At shifts smaller than 0.4 pixels, for a SNR < 5, the CEE remained greater than 100%. The performance of CoG at shifts of 0.4, 0.8, 1.5 pixels is shown in Fig. 4. Clearly CoG cannot be used in the case of Shack Hartmann sensor. This is because a Shack Hartmann sensor makes differential measurements of wave-fronts that differ from each other by a very small phase correspondingly giving minor shifts.

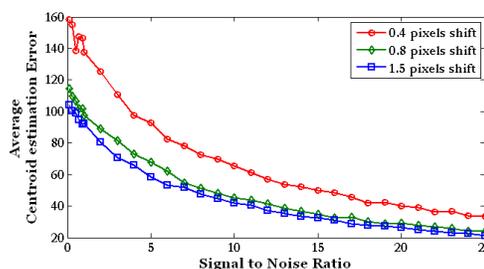

Fig. 4 CoG: Varying shift of the spots

402

2) *IWCoG:* Shifts bigger than 0.1 pixels can be easily detected with an error of 0.05 pixels when SNR > 1. Shifts smaller than 0.1 can be detected only with a better SNR as shown in Fig. 5.

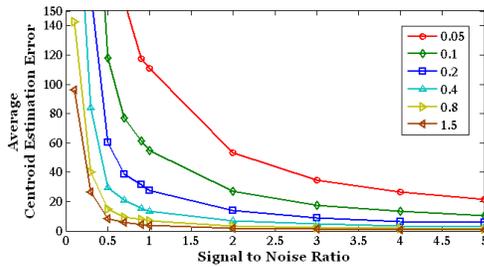

Fig. 5 IWCoG: Varying shift of the spots

3) *IWC:* The behaviour is similar to IWCoG with a difference that it requires even better SNR for low shifts as shown in Fig. 6.

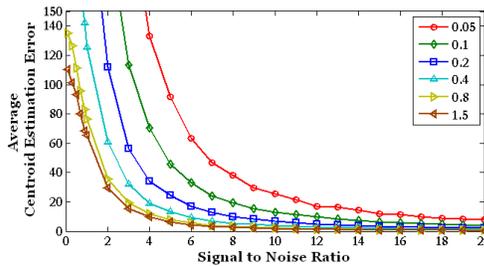

Fig. 6 IWC: Varying shift of the spots

## C. Dependence on spot size:

The size of the spots of a typical Shack Hartmann varies from 10-50μm when exposed to sufficient amount of light. The spot size in the simulations was characterized in terms of the FWHM of the Gaussian spot. We used spots with FWHM=2.5, 3.5, 4.5 pixels.

In the case of CoG, a larger spot reduces CEE compared to a smaller spot as shown in Fig. 7. This graph shows a comparison of the spot size and the CEE when the shift of 1.5 pixels is simulated.

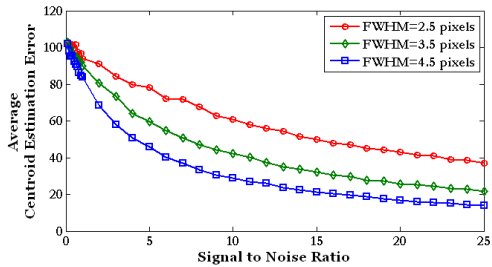

Fig. 7 Performance of CoG with changing spot size at 1.5 pixel shift

In the case of IWCoG too, the CEE reduces with increasing size of the spot for SNR < 0.5 as shown in Fig. 8. This is because, in the case of smaller spots, fitting a Gaussian leads to discretization errors and the presence of noise distorts the edges of the spot. Above SNR=0.5, CEE remains same for any of the spot sizes.

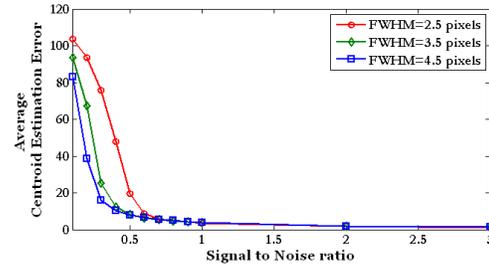

Fig. 8 Performance of IWCoG with changing spot size at 1.5 pixel shift

IWC also shows a similar behaviour as shown in Fig. 9. CEE reduces with spot size and it remains same for SNR > 10.

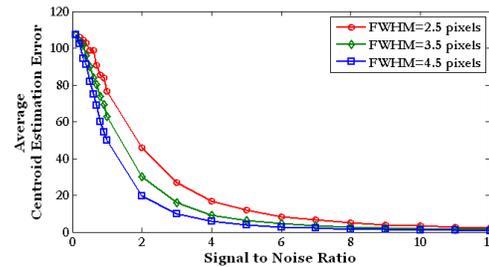

Fig. 9 Performance of IWC with changing spot size at 1.5 pixel shift

## D. Error Propagation in IWCoG

Any iterative problem needs an optimization process. In the case of iterative weighting also, we need to optimize the number of iterations required to minimize the error in estimating the actual centroid position. Iteration optimization is required for two reasons, firstly to remove redundancy of computing excessive iterations without any improvement and secondly to pick up exactly the number of iterations that leads to minimum CEE.

403

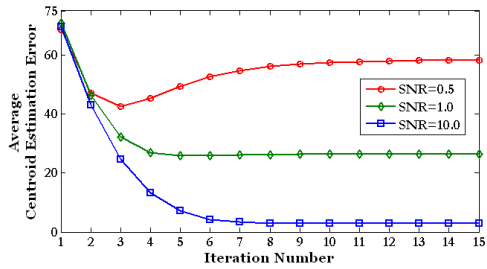

Fig. 10 Error Propagation in the iterative process of IWCoG method

Fig. 10 shows for a particular case of 0.2 pixels shift in a spot with 2.5 pixels FWHM. For a SNR of 0.5, there is dip at third iteration in CEE. When SNR is greater than 1, the centroiding error reduces to a minimum and then maintains that value with increasing iteration number. From Fig. 10 we can conclude that the optimum number of iterations for SNR=0.5, 1 and 10 in this case is 3, 5 and 8 respectively. More investigation on the optimum number of iterations should be done before applying this technique in real systems.

### E. *Weighting using higher powers of intensity*

For an image, the centroid can be obtained most accurately by using the intensity as the weighting function when we do not have any information about the shape of the object. If we start with a most general expression for the centroid estimation error and minimise the error with respect to the weighting function, we will arrive at the intensity function. Using higher powers of intensity as weighting function only means that we are trying to pick up the point of maximum intensity from a sample by raising the intensity of the image to higher powers. The results of using intensity raised to P=1, 2, 3, 6 as weighting function for shift of 0.2 is shown in Fig. 11. Fig. 12 shows a comparison in the case of 7 pixel shift in the spot.

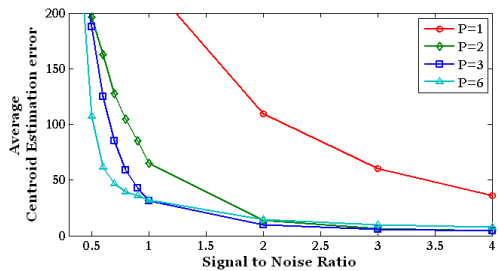

Fig. 11 Higher powers of intensity as weighting function, 0.2 pixel shift

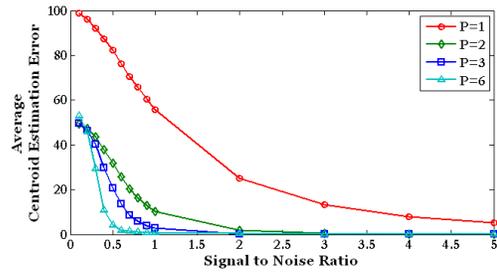

Fig. 12 Higher powers of intensity as weighting function, 7 pixels shift

### F. *Error Stability*

Since the Shack Hartmann Sensor measures a large set of slopes in terms of shifts in the spots and then uses these slope values for construction of the phase function, the centroiding error at a particular noise level has to maintain a stable value independent of the magnitude of the shift in the spot. The error stability will allow us to calculate the error in estimation of the wave-front better. The error is more stable in the case of IWCoG compared to CoG and IWC as examined earlier in Figs. 4, 5, 6.

### G. *Computational Speed*

Speed of computation is critical when we use high resolution sensors. Clearly CoG takes minimum time for computation compared to IWCoG and IWC. Computations were done using a 3 GHz Pentium IV PC with 1 GB RAM. The code is designed using MATLAB R2008a. For a 50×50 pixels image, CoG took 85µs to estimate the centroid position. The time taken for IWCoG for a similar image with increasing number of iterations is shown in Fig. 13.

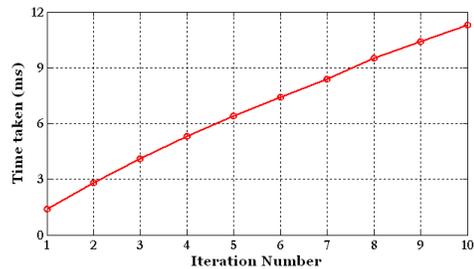

Fig. 13 Time taken to compute IWCoG as a function of iteration number

The computational speed of IWC depends on the power of intensity used and is 4.8 ms for P=5; and 0.1ms for P=2. We need smarter techniques for efficient and faster centroiding, since there are many applications like adaptive optics where we cannot make a compromise on quality or speed.

### H. Temporal Stability

To test the stability of centroiding, a lab experiment was performed where a collimated beam was aberrated using a CD case as shown in Fig. 14. A 15mW He-Ne laser was used as the source. G.P is a glan polarizer and S.F represents a spatial filter setup. The beam was collimated using a doublet, L1 of focal length 20cm. An array of Diffractive Optical Lenses (DOLs) with a focal length of 13.5cm was projected on the SLM, which acts like a Shack Hartmann Sensor. The image at the focal plane of the array was reimaged onto the CCD (Pulnix) after demagnification using lenses L2 and L3.

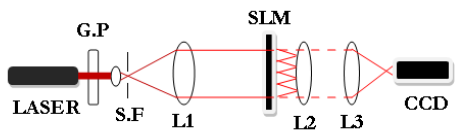

Fig. 14 Experimental setup to measure the centroiding stability

Images of the Shack Hartmann Sensor (Spatial Light Modulator based) spot pattern were recorded using a CCD placed at the focal plane with a delay of 1s without changing any experimental parameter. The background intensity distribution was found to be following Gaussian statistics. The images recorded were used for stability analysis. The position of the centroid was estimated using the CoG, IWCoG and IWC. The standard deviation of the estimated centroid position was smaller in the case of IWCoG (0.03 pixels) as compared to IWC (0.06 pixels) and CoG (0.27 pixels).

### IV. CONCLUSIONS

Normalized Center of Gravity (CoG), Iteratively Weighted Centre of Gravity (IWCoG) and Intensity Weighted (IWC) centroiding methods were used to compare the centroiding accuracy in the presence of uniform background noise in the case of a Shack Hartmann sensor. We indicate the importance of the optimization of the performance of iterations of IWCoG method. The optimum number of iterations is dependent on the signal to noise ratio. The use of higher powers of intensity as weighting function at different noise levels was analysed. It was found that the computational efficiency can be improved manyfold by understanding the performance of these algorithms.

ACKNOWLEDGMENT

We wish to thank Avinash A. Deshpande for his useful suggestions.

**References**

[1] D. G. MacMartin, "Control challenges for extremely large telescopes," in Smart Structures and Materials 2003: Industrial and Commercial Applications of Smart Structures Technology, E. H. Anderson, ed., *Proc. SPIE 5054*, 275–286 (2003).

[2] Daniel R. Neal, James Copland, and David A. Neal, "Shack-Hartmann wavefront sensor precision and accuracy," *Proc. SPIE 4779*, 148 (2002).

[3] S. Thomas, T. Fusco, A. Tokovinin, M. Nicolle, V. Michau, G. Rousset, "Comparison of centroid computation algorithms in a Shack Hartmann sensor", *MNRAS*, 371, 1, 323 (2006).

[4] K. L. Baker and M. M. Moallem, "Iteratively weighted centroiding for Shack- Hartmann wave-front sensors," Opt. Express **15**, 5147-5159 (2007).